# Flattening the COVID-19 Curve: The "Greek" case in the Global Pandemic

**Konstantinos Demertzis[1]\*, Lykourgos Magafas[1] and Dimitrios Tsiotas[2,3]**

[1] Laboratory of Complex Systems, Department of Physics, Faculty of Sciences, International Hellenic University, Kavala Campus, St. Loukas, 65404, Greece; kdemertzis@teiemt.gr; magafas@teikav.edu.gr;

[2] Department of Regional and Economic Development, Agricultural University of Athens, Greece, Nea Poli, Amfissa, 33100, Greece; tsiotas@aua.gr;

[3] Department of Planning and Regional Development, University of Thessaly, Pedion Areos, Volos, 38334, Greece; tsiotas@uth.gr;

\* Correspondence: kdemertzis@teiemt.gr;

**Abstract:** The global crisis caused by the COVID-19 pandemic, in conjunction with the economic consequences and the collapse of health systems, has raised serious concerns in Europe, which is the most affected continent by the pandemic since it recorded 2,388,694 cases and 190,091 deaths (39.6% of the worldwide total), of which 71.7% (136,238) are in the United Kingdom (43,414), Italy (34,708), France (29,778), and Spain (28,338). Unlike other countries, Greece, with about 310 confirmed cases and 18 deaths per million, is one bright exception in the study and analysis of this phenomenon. Focusing on the peculiarities of the disease spreading in Greece, both in epidemiological and in implementation terms, this paper applies an exploratory analysis of COVID-19 temporal spread in Greece and proposes a methodological approach for the modeling and prediction of the disease based on the Regression Splines algorithm and the change rate of the total infections. Also, it proposes a hybrid spline regression and complex network model of social distance measures evaluating and interpreting the spread of the disease. The overall approach contributes to decision making and support of the public health system and to the fight against the pandemic.

## 1. Introduction

The SARS-CoV-2, which causes the COVID-19 disease, is responsible for the evolving coronavirus pandemic of the period 2019-20, causing a state of emergency of the public health system (Ahmed et al., 2020). The pandemic cannot be prevented from transmission between persons since no vaccine currently exists and thus it inevitably infects millions of people around the world. In particular, more than 9.9 million people have been reported to be infected by the new coronavirus worldwide, and nearly 497,000 have died so far (Roser and Ritchie, 2020; source: https://www.worldometers.info/, accessed: 30/6/20). Since the emergence of the disease in China, in December 2019, more than 215 countries and regions have reported infected cases (Roser and Ritchie, 2020). The COVID-19 has a single-strand positive RNA genome of positive polarity. Like most RNA viruses, SARS-CoV-2 shows strong variability in its genetic material and a tendency to mutate. It is the seventh coronavirus along with the SARS-CoV (SARS disease), MERS-CoV (MERS disease), HKU1, NL63, OC43, and the 229E, which can infect humans (Lescure et al., 2020). Exposure to SARS-CoV and MERS coronaviruses can induce immunity for about 2 (maybe 3 in the case of MERS) years, which then resolves. Typically, SARS-CoV-2 is believed to cause immunity for approximately 1 year (similarly to the seasonal coronaviruses), although questions about the duration of immunity for the asymptomatic patients or for patients with mild symptoms have not been answered yet (Ahmed et al., 2020; Xu et al., 2020).

The COVID-19 pandemic is considered to be zoonotic and is closely related to (79.5%) the original SARS-CoV (Ahmed et al., 2020). Genetic analysis has revealed that it is related to the genus Betacoronavirus, in the B series of the Sarbecovirus subgenus, along with two components derived from the bat (Lescure et al., 2020; Xu et al., 2020). Its genome is 96% identical to other coronavirus samples (BatCov RaTG13), while Chinese researchers have found that it differs only to one amino-acid in genome sequences between viruses found in the anteater Pagolins and those derived from humans (Lam et al., 2020), implying that Pagolins may have been an intermediate host.

The virus is transmitted from person to person through respiratory droplets produced during coughing. The day of exposure to the onset of symptoms (incubation time) varies from 2 to 14 days, but it usually refers to 5 days (Fang et al., 2020; Heymann and Shindo, 2020). Patients infected by the virus may be asymptomatic or have symptoms similar to the common cold (i.e. fever), cough, and dyspnea. A key factor of SARS-CoV-2 high transmissibility is the high levels of multiplication of this virus in the upper respiratory region, even before the symptoms appear (Ahmed et al., 2020; Fang et al., 2020). The virus is also transmitted by touching a surface or an object, where the virus is located, and then eyes, nose, or mouth are touched (Bai et al., 2020; Heymann and Shindo, 2020).

Samples taken for molecular detection of SARS-CoV-2 are mainly either oral pharyngeal or nasopharyngeal coating, or pharyngeal washing from the upper respiratory region, or bronchial excretes from the lower respiratory region. Clinical samples taken from the lower respiratory region are preferable in terms of their diagnostic value than those from the upper respiratory region (Bai et al., 2020). The sample is taken by using a special stick (plastic stick with a dacron top) in the case of coating sampling or by a sterile bottle in all other cases. After the samples are received by the laboratory, the viral RNA isolation procedure follows, where the reverse transcription polymerase chain reaction is applied to amplify 2 different targets of the SARS-CoV-2 viral genome. The test is completed within 3 to 4 hours of receiving the samples from the laboratory (La Marca et al., 2020).

A negative result for SARS-CoV-2 implies that the virus is not detected in the sample. Except from the uninfected individuals, a negative result can also occur in asymptomatic carriers running the incubation phase. The molecular method outperforms all other methods because is able to detect the virus in the very early stages of the infection, but it cannot definitely (100%) detect the virus in the incubation phase (Corman et al., 2020). Also, it is very important in receiving the coatings to have enough collected material available. In general, the nasopharyngeal coating is better than the oropharyngeal one, since the collected material is usually more in quantity and is unaffected by the recipient or the patient. In symptomatic patients the sensitivity and specificity of the method is over 99% (Rosado et al., 2020).

There is currently no certified vaccine (although the production process is pending) or effective treatments for COVID-19 and therefore all efforts are restricted to symptom management and to supportive measures, such as providing oxygen, monitoring of vital organ function, and providing intravenous fluids, where needed (Ahmed et al., 2020). However, more than 70 drug substances and their combinations are already under clinical research and their results will be gradually known (Beigel et al., 2020). The most significant developments up to date concern the intravenous antiviral remedisivir, which is broad-spectrum and inhibits the proliferation of viruses that have RNA genetic material, such as the coronavirus. Also, chloroquine phosphate appeared able to reduce the duration of infection and the days of viremia and it improved the lung function and the outcome of pneumonia relatively to placeboes, in a sample of patients (Gao et al., 2020).

Due to the lack of vaccine and antiviral drugs so far, the most effective way to fight the disease is still, from the side of the society, the implementation of social distancing measures, the collective activation, and the individual and social responsibility of each citizen, while, from the side of the scientific research, the development of appropriate methods for further understanding and detection of the disease, as well as the improvement of health resources management (Tsiotas and Magafas, 2020).

It should be noted that the study of the pandemic exclusively by epidemiological methods creates serious concerns about the quality of research and its expected results. This is because an epidemiological model uses the microscopic description (i.e. builds on the information provided by an infected person) to predict the macroscopic behavior of the disease spreading across the population (Demertzis et al., 2020). Quantitative predictions of epidemiological methods also involve uncertainty, as the models are usually theoretical with many simplifications and assumptions, where many parameter values can only be estimated and not accurately measured (Schlickeiser and Schlickeiser, 2020). Another serious drawback in epidemiological research is that the available data are usually insufficient due to the lack of information on the range of values in some parameters, while experiments on infectious diseases in human populations are either impossible or immoral to implement. The lack of reliable information and the impossibility of applying repetitive experiments necessitate the use of mathematical modeling in the study and prediction of the COVID-19 spread.

Within this context, the interpretation, modeling, analysis, and prediction of the temporal spread of the disease is extremely important, not only from an epidemiological point of view, but mainly from a mathematical modeling perspective. Also, it is very important to develop models analyzing the epidemiological phenomena of the disease, in order to predict future conditions and especially the variation of the spread curve of the disease and its time horizon of normalization. Within this context, this paper applies an exploratory analysis of COVID-19 temporal spread in Greece and proposes a methodological approach for the modeling and prediction of the disease based on the Regression Splines algorithm and the change rate of the total infections. Also, it proposes a hybrid complex network model of social distancing measures evaluating and interpreting the spread of the disease. The overall approach contributes to decision making and support of the public health system and to the fight against the pandemic. The remainder of this paper is organized as follows; Section 2 reviews the history of COVID-19 in Greece, Section 3 reviews the current methods for modeling and predicting the temporal spread of the disease, Section 4 presents the methodology and data, Section 5 applies the proposed methodology on the dataset of Greek COVID-19 infection curve, and Sections 6 and 7 are the discussion and conclusions sections.

## 2. Flattening of the Curve

The infection mechanism of COVID-19 between individuals is already known almost from the first days of the emergence of the pandemic (Ahmed et al., 2020; Bai et al., 2020; Heymann et al., 2020). On the contrary, the social spreading of COVID-19 within a population is a quite complex procedure and therefore it is very difficult to understand its dynamics without employing mathematical modeling and complexity methods. Although their default limitations, mathematical models are very effective in epidemiology because the can go beyond normality or randomness and describe more complicate structures (Tsiotas and Magafas, 2020). Usually, the evolution of the pandemic can be shown in a 2D diagram, where the $x$-axis represents time (usually days) and the $y$-axis represents either the additive or the cumulative infections data. As it shown in Fig.1, the initial increase of the COVID-19 infection curve appears to be exponential, since an abrupt increase is recorded on 10 Mar and afterwards.

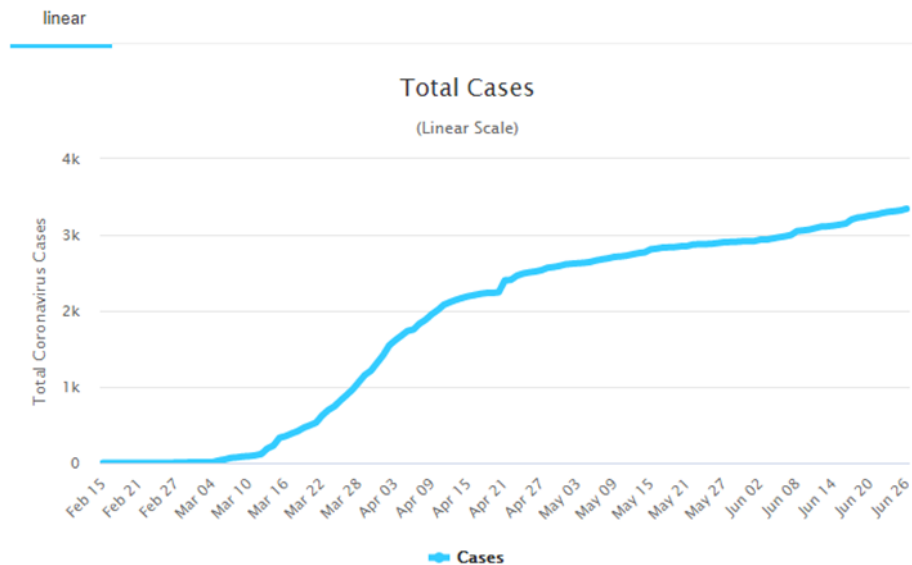

**Fig.1.** Line-plot showing the Greek COVID-19 infection curve, for the period 15 Feb 2020 up to 26 June 2020 (source: https://www.worldometers.info/, accessed: 30/6/20)

According to Fig.1, the temporal spread of the Greek COVID-19 illustrates a stochastic system with possibly some linear correlations of short duration. Generally, the epidemiologic aim for the spread of a disease is shorten the height of and to lengthen the width of the curve (i.e. the increase of cases to be smooth and not abrupt), while an important forecasting issue is to estimate the peak of the spreading, where the decline stage (flattening of the curve) then begins (Schuttler et al., 2020).

The flattening of the infection curve is crucial for how effectively a national health system will respond and therefore how many lives will eventually be saved. Specifically, the transmission rate of the infection and the consequent increase of the infections will determine the mortality rate of COVID-19 (Schlickeiser and Schlickeiser, 2020). A high slope of the epidemic curve may imply a deficiency either in intensive care unit (ICU) beds, or in medical or nursing staff, or in medical equipment and supplies (e.g. respirators, etc.). Accordingly, the flatter the epidemic curve over time is, the fewer people will need hospitalization at any given time (Demertzis et al., 2020). Within this context, mathematical modeling of the disease spread and particularly the forecasting methods of the evolution of the epidemic curve and of its flattening are a constant demand of research for the academic community, while remarkable findings have already been recorded offering an important inventory for the fight against the disease.

## 3. Related Work

A typical example of a method estimating the duration of infection and the expected number of infections is presented in the work of Barmparis and Tsironis (2020). This paper studies the infection curves from eight countries with reference to the outbreak of disease in China, aiming to estimate the evolution of the infection, the expected number of daily infections per country, and (perhaps the most important) the duration of the epidemic in each of these countries. The analysis showed that Italy, Spain, and the Netherlands have already passed the peak-point of the disease, while Greece, France, and Germany were close to it. Also, the work of Ranjan (2020) estimates the spreading of COVID-19 along different geographical areas by using the logistic, the SIR, and generalized SEIR models. The analysis showed that both the SIR and the generalized SEIR models yield similar estimates for the areas where epidemic curves show flattening trends. According to these models, the final size of the epidemic in the US, Italy, Spain, and Germany could be 1.1, 0.22, 0.24, and 0.19 million infected cases respectively. Further, the purpose of the work of Kiesha et al. (2020) is to calculate the effects of the social distance measures in the evolution of the COVID-19 pandemic and therefore to provide useful insights and practices to other countries.

On the other hand, provided that a significant number of infections (including human coronaviruses) follow seasonal patterns, the study of Sajadi et al. (2020) examined ERA-5 climate data, from cities with significant COVID-19 spread, in comparison with those extracted from areas that have not yet been affected or the infection has not significantly spread. Finally, the work of Petropoulos and Makridakis (2020) tries to objectively predict the evolution of COVID-19 by using a simple but robust methodology. The authors assume that the available infection data, the number of deaths, and recoveries are reliable and that the future spreading of the disease will follow the same pattern of the past. This work suggests a good example in predicting the temporal spread of the disease, having potentially large implications in terms of planning and decision making.

Whether focusing on the peculiarities of the Greek COVID-19 spread, both epidemiologically and in terms of policy implementation, this paper provides an exploratory temporal study based on the analysis of COVID-19 infection data in Greece.

## 4. Dataset and Methodology

To accurately approach the modeling problem, the prime goal is to find the mathematical expression that can model the data of the disease spread in Greece and to describe how cases increase over time. At next, the rate of change of the disease is calculated and the variation of the available variables is approached with the Regression Splines methodology, which provides the best possible fitting to the available data (Norusis, 2008; De Boor et al., 2020). Finally, a novel method that was proposed by Demertzis et al. (2020) for predicting the flattening of the curve is implemented, which is based on a hybrid Regression Splines and Complex Network approach.

### 4.1. Data

The available data were drafted from the Hellenic Ministry of Health and are freely available on the official website https://eody.gov.gr/ of the Ministry in Greece. The data include daily measurements during the period from 26 Feb 2020 to 26 June 2020 and regard the day of cases (variable: Day_ID), the total infected cases (variable: All_Cases), the daily new cases (variable: New_Cases), the daily new deaths (variable: New_Deaths), daily recovered cases (variable: Recovered), daily patients entering to ICUs (variable: ICU), the daily change of active cases (variable: Active_Cases), and the diagnostic tests performed per day (variable: Tests).

A first observation to the measurements related to the COVID-19 spread in Greece shows that this dataset is a continuous time-series, which initially appears inclining trend and afterwards shows signs of stability. Also, no time-variant fluctuations are observed, since the time-series does not show periodicity (seasonality) or cyclic structure. Although the available sample is not sufficiently enough, these observations illustrate that the COVID-19 temporal spread in Greece is more closely to a structure of stationary than a periodic structure. A more in-depth analysis builds on understanding the historical behavior of the disease and on estimating the parameters describing a good model fitting to the available time-series data.

*4.2. Regression Splines*

Regression splines are parts of a segmented polynomial function that maintains a high smoothness at the points connecting successive polynomials (De Boor et al., 2020). They are essentially segmented polynomial approaches, which are very useful due to their flexibility, simplicity, ease, accuracy of evaluation, and their ability to approximate complex curves by fitting simple polynomial models (Norusis, 2008).

As previously shown, while polynomial regression can be very effective in terms of estimation, it is complex in deciding the polynomial degree and therefore the type of model that better fits to the available data (Walpole et al., 2020). For instance, choosing a high degree polynomial may cause over-fitting issues, while fitting a low degree polynomial usually leads to loss of information. Also, the polynomial degree is directly related to the bias-variance tradeoff (Geman et al., 2020). That is, a high-order polynomial is directly related to high variance, while a low-order polynomial describing the same model has low variance. On the other hand, high-order polynomials have high determination and therefore low bias, while low-order polynomials have low determination that leads to high bias. This bias-variance tradeoff is described by the expression (Demertzis et al., 2020):

$$expected_{loss} = (bias)^2 + variance + noise \tag{1}$$

where

$$(bias)^2 = \int \{E_D[y(x; D)] - h(x)\}^2 p(x)dx$$

$$variance = \int E_D[\{y(x; D) - E_D[y(x; D)]\}^2] p(x)dx$$

$$noise = \iint \{h(x) - t\}^2 p(x, t)dxdt$$

In general, simple models are described by small variance but high determination, while more complex models by small loss of information but high variability (Geman et al., 2020), a fact that validates the bias-variance trade-off rule. Within this context, the choice for the best model requires a balance between level of determination and bias (Demertzis et al., 2020). Also in cases of variant volatility, parts of non-linear behavior, temporary randomness, and generally irregular structure of the time-series data, the polynomial regression model will not provide good fittings. This is the main reason why polynomial regression is suitable for modeling smoothed data (Norusis, 2008; Walpole et al., 2020). Overall, provided that they yield high determination, polynomial regression models should be used in typical cases to avoid excessive solutions leading to over-fitting.

In contrast, splines regression applies low-degree polynomial fittings, avoiding thus the Runge effect (Stone et al., 1997) describing higher-order polynomial approaches. The rationale of regression splines is based on the fact that every polynomial function can be written as a linear combination of simpler functions (Geman et al., 2020). The range of the independent variable is divided into "knots", which define the end of one spline part and the beginning of the next. Overall, the splines are defined so that the resulting fitting curve to be smooth and continuous, thus limiting the variable to be linear at the edges (Demertzis et al., 2020). For instance, the expression:

$$\{h_k(x) - t\}_{k=1,\dots,J} : r(x) = \sum_{k=1}^{J} \beta_k h_k(x) \qquad (2)$$

is defined by the terms $h_k(x)$ that may express $m-$degree polynomials or a summand of linear functions defined in the domain $\{A_k = [\xi_k, \xi_{k+1}]\}_{k=1,\dots,K}$ according to the expression (Stone et al., 1997; Geman et al., 2020):

$$h_k(x) = I_{A_k}(x) \text{ and } h_{\frac{k}{2+k}}(x) = I_{A_k}(x), \text{ for } k = 1,\dots,\frac{k}{2}, \text{ thus } K = J \qquad (3)$$

Creating a spline requires defining knots on the spline curve so that to be as smooth as possible. Two choices are here required; the first is to define the number and position of knots and the second to define the degree of polynomial fitted to the interval defined between successive nodes (Demertzis et al., 2020). The major demand is to define individual curves so that to intersect in a "smoothly" way. Therefore, a spline function is a group of "smoothly connected" square polynomials, implying that, for polynomials of degree m, both the spline function and their first derivatives m-1 should be continuous at the knots. In case that $k$ knots are used, fitting a polynomial of degree $m$ requires estimating $k+m-1$ regression parameters.

For instance, for polynomials of degree (M–1), which are defined in K intervals and have their first (M–2) derivatives continuous at the boundaries of the intervals (nodes), the spline fitting is calculated as follows (Geman et al., 2020):

$$h_1(x) = 1, h_2(x) = x, h_M(x) = x^{M-1}, h_{M-1+k}(x) = (x - \xi_k)_+^{M-1} \quad k = 2,\dots,K \qquad (4),$$

or equivalently:

$$J = (M) + [K - 2 + 1] = M + K - 1 \qquad (5).$$

Either a few in number intervals $K$ or low polynomial-order $M-1$ implies high bias, whereas the opposite expresses σημαίνει high variance.

It has been empirically shown that five knots suffice to create several non-linear paths (Demertzis et al., 2020). In practice, cubic splines (i.e. third-order polynomials) are commonly used, which allow modeling the curve structures and simultaneously provide sufficient fitting flexibility, whereas they do not require as many degrees of freedom as higher-order splines do (Geman et al., 2020; Demertzis et al., 2020). To estimate the regression splines, we define the X matrix as follows (Geman et al., 2020):

$$X = \begin{bmatrix} h_1(x_1) & \cdots & h_J(x_1) \\ \vdots & & \vdots \\ h_1(x_n) & \cdots & h_J(x_n) \end{bmatrix} \qquad (6),$$

and the interpolation matrix L as follows (Geman et al., 2020):

$$L = X(X^T X)^{-1} X^T \qquad (7).$$

Within this context, the least square estimators are defined by the relation (Geman et al., 2020):

$$\hat{r}(x_i) = \sum_{j=1}^{J} z_{ij} y_j : \underline{\hat{r}} = L\underline{y} \qquad (8),$$

which yields (Geman et al., 2020):

$$E(\underline{\hat{r}}) = L \cdot \underline{\hat{r}} \qquad (9),$$

where $\underline{\hat{r}} = (r(x_1) \dots r(x_n))^T$, $var(\underline{\hat{r}}) = \sigma^2 L L^T$, which equals to $var[\hat{r}(x_i)] = \sigma^2 \|\underline{z}(x_i)\|^2$ with $\underline{z}(x_i)$ defined by the $z$-th row of $L$ as follows (Geman et al., 2020):

$$CV = \frac{1}{n} \sum \left( \frac{y_i - \hat{y}_i}{1 - z_{ii}} \right)^2 \qquad (10).$$

Although Regression Splines suggest an excellent fitting method for modeling the temporal spread of COVID-19 in Greece (a method that can provide solutions in several malfunctions related to randomness and generally to non-linear structures of the time-series), the major problem of this method is its dependence on the usually arbitrary choice of knot defining the spline parts of the overall curve (Demertzis et al., 2020). In the literature (Traa and Smaragdis, 2014; Zhou et al. 2018; Chen et al. 2019), most of the repairing approaches rely on purely heuristic conceptualization building either on visual differences, or seasonal behavior, or user experience, or experts' opinions.

The criterion used in this paper to define the spline knots is based on a recent methodology proposed by the authors of this paper (Tsiotas and Magafas, 2020; Demertzis et al., 2020), which builds on complex network analysis to evaluate the effectiveness of the COVID-19 temporal spread in Greece. The analysis revealed five distinct parts in the structure of the COVID-19 infection curve, which are defined by the intervals $Q_1 = [1-4] \cup [9-19]$, $Q_2 = [5-8]$, $Q_3 = [20-26]$, $Q_4 = [27-32]$, and $Q_5 = [33-43]$ (Tsiotas and Magafas, 2020). In this outcome, the definition of the non-convex $Q_1$ interval, which is impossible to detect by other either linear of non-linear time-series approaches, illustrates the added value of the proposed methodology.

## 5. Forecasting

To predict the COVID-19 temporal spread in Greece within the best possible determination ability, we apply a combined model of complex-network defined splines, as proposed by Demertzis et al (2020). In particular, on the available set of time-series data $(x_1, f(x_1)), ..., (x_n, f(x_n))$, the proposed model calculates an optimal approximation of the pairwise polynomial function defined on the sub-domains $Q_1$, $Q_2$, $Q_3$, $Q_4$, and $Q_5$ that are computed by the complex-network approach proposed by Tsiotas and Magafas (2020). The main goal of the approach is to construct the best possible fitting model describing the Greek COVID-19 infection curve.

### 5.1 Rate of Change

Prior forecasting, the change rate of the COVID-19 spread in Greece should be calculated. By considering the increase of infected cases as a sequence, we can define two successive terms $a_v$ and $a_{v+1}$ with the relation $a_{v+1} = a_v + m$, where $m$ is the change rate of the sequence. Within this context, the relation between the change rate $m$ and the total number of daily infection is expressed as follows:

$$m = \frac{\Delta y_t}{\Delta y_{t+1}} \tag{11},$$

where $\Delta y_t$ is the number of new cases in day $t$ and $\Delta y_{t+1}$ in day $t+1$. That is, a unit change of the quantity $\Delta y_1$ induces an $m$-times change of $\Delta y_2$. When $m>0$, the sequence is ascending, when $m<0$ the sequence is decreasing, and when $m=1$ the sequence is constant. The change rate of the available 65-day period is shown in Fig.2.

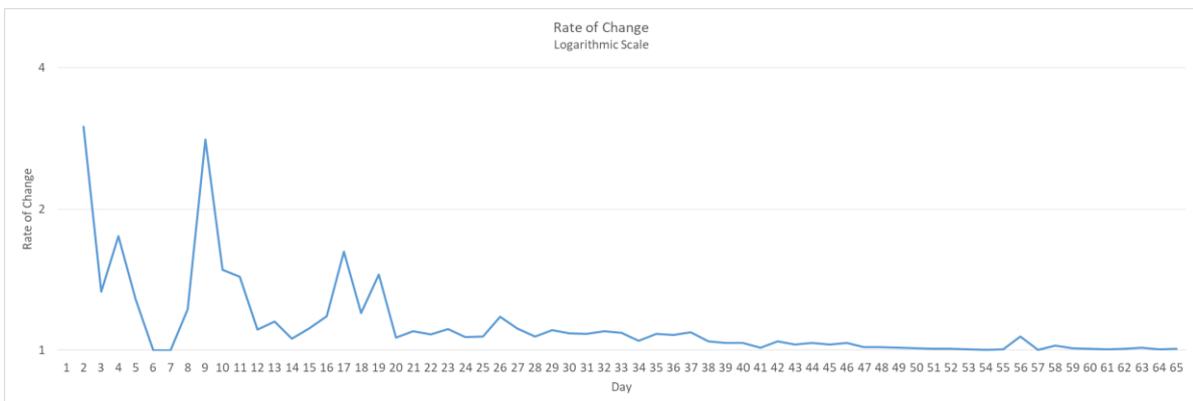

**Figure 2.** Line plot of the Change Rates of all cases in Greece (https://eody.gov.gr/)

As it can be observed, there were significant fluctuations in the change rate in the first 19 days, while at next the rate follows a smooth trend. The average change rate was calculated according to the equation:

$$\langle m \rangle = \frac{1}{n} \sum_{i=1}^{n} y_i = \frac{1}{n}(y_1 + \cdots + y_n) \tag{12}.$$

For $n$ = 46 (which refers to the period of 46 days being studied that starts from 19th day prior to which the change rate includes significant fluctuations) the average change rate yields $\langle m \rangle$=1.049521.

*5.2 Regression Splines Forecasting*

The modeling approach based on Regression Splines builds on the measure of change rate defined in relation (11). Provided that the time-series being under consideration has a constant change rate, the Regression Splines Algorithm (RSA) was used to forecast a period of the next 62 days (see Table 1), until 01 July 2020, which refers to the date of tourism opening starts in Greece.

**Table 1.** Regression Splines Forecasting

| ID | Day | Day_ID | Regression Splines Forecasting |
|----|-----|--------|-------------------------------|
| 1. | 1-May-20 | 66 | 2602 |
| 2. | 2-May-20 | 67 | 2614 |
| 3. | 3-May-20 | 68 | 2626 |
| 4. | 4-May-20 | 69 | 2638 |
| 5. | 5-May-20 | 70 | 2650 |
| 6. | 6-May-20 | 71 | 2662 |
| 7. | 7-May-20 | 72 | 2674 |
| 8. | 8-May-20 | 73 | 2687 |
| 9. | 9-May-20 | 74 | 2699 |
| 10. | 10-May-20 | 75 | 2711 |
| 11. | 11-May-20 | 76 | 2724 |
| 12. | 12-May-20 | 77 | 2736 |
| 13. | 13-May-20 | 78 | 2749 |
| 14. | 14-May-20 | 79 | 2761 |
| 15. | 15-May-20 | 80 | 2774 |
| 16. | 16-May-20 | 81 | 2786 |
| 17. | 17-May-20 | 82 | 2799 |
| 18. | 18-May-20 | 83 | 2812 |
| 19. | 19-May-20 | 84 | 2825 |
| 20. | 20-May-20 | 85 | 2838 |
| 21. | 21-May-20 | 86 | 2850 |
| 22. | 22-May-20 | 87 | 2863 |
| 23. | 23-May-20 | 88 | 2877 |
| 24. | 24-May-20 | 89 | 2890 |
| 25. | 25-May-20 | 90 | 2903 |
| 26. | 26-May-20 | 91 | 2916 |
| 27. | 27-May-20 | 92 | 2929 |
| 28. | 28-May-20 | 93 | 2943 |
| 29. | 29-May-20 | 94 | 2956 |
| 30. | 30-May-20 | 95 | 2970 |
| 31. | 31-May-20 | 96 | 2983 |
| 32. | 1-Jun-20 | 97 | 2997 |
| 33. | 2-Jun-20 | 98 | 3011 |
| 34. | 3-Jun-20 | 99 | 3024 |
| 35. | 4-Jun-20 | 100 | 3038 |
| 36. | 5-Jun-20 | 101 | 3052 |
| 37. | 6-Jun-20 | 102 | 3066 |
| 38. | 7-Jun-20 | 103 | 3080 |
| 39. | 8-Jun-20 | 104 | 3094 |
| 40. | 9-Jun-20 | 105 | 3108 |
| 41. | 10-Jun-20 | 106 | 3122 |

| ID | Day | Day_ID | Regression Splines Forecasting |
|---|---|---|---|
| 42. | 11-Jun-20 | 107 | 3136 |
| 43. | 12-Jun-20 | 108 | 3151 |
| 44. | 13-Jun-20 | 109 | 3165 |
| 45. | 14-Jun-20 | 110 | 3180 |
| 46. | 15-Jun-20 | 111 | 3194 |
| 47. | 16-Jun-20 | 112 | 3209 |
| 48. | 17-Jun-20 | 113 | 3223 |
| 49. | 18-Jun-20 | 114 | 3238 |
| 50. | 19-Jun-20 | 115 | 3253 |
| 51. | 20-Jun-20 | 116 | 3268 |
| 52. | 21-Jun-20 | 117 | 3283 |
| 53. | 22-Jun-20 | 118 | 3298 |
| 54. | 23-Jun-20 | 119 | 3313 |
| 55. | 24-Jun-20 | 120 | 3328 |
| 56. | 25-Jun-20 | 121 | 3343 |
| 57. | 26-Jun-20 | 122 | 3358 |
| 58. | 27-Jun-20 | 123 | 3373 |
| 59. | 28-Jun-20 | 124 | 3389 |
| 60. | 29-Jun-20 | 125 | 3404 |
| 61. | 30-Jun-20 | 126 | 3420 |
| 62. | 1-Jul-20 | 127 | 3435 |

The RSA was run under the error minimization criterion (Stone et al., 1997; Geman et al., 2020). In particular, the model was constructed according to the following relations:

$$\hat{f}(x_k) = h(x_k) - g(x_k)$$
$$g(x_k) = \left(\hat{f}(x_{k-1}) * U(x_k)\right) * \langle m \rangle$$
$$U(x_k) = \langle m \rangle - U(x_{k-1})$$
$$h(x_k) = (g(x_k) * m) + \left(g(x_k) * U(x_k)\right) * \langle m \rangle$$

(13),

where $\langle m \rangle$ is the average change rate, $\hat{f}(x_{k-1})$ is the number of infection in day $k$-1, $U(x_k)$ is the daily change rate for day $k$, which is defined by the residual of the average change rate minus the daily change rate of day $k$-1. As it come of from this optimization approach, the prediction error is significantly smoothened due to the polynomial nature of the Regression Splines algorithm, which increases exponentially for long-term predictions. Accordingly, the change rate remains stable, which verifies the effectiveness of the methodological approach.

*5.3 Flattening of the Curve*

Based on the previous approaches, we develop a combined method for flattening the COVID-19 infection curve in Greece. Provided that the problem is initially described by a stage of exponential growth, in which the growth rate gradually decreases and ends asymptotically to growth saturation, we apply a Logistic regression prediction technique, with sigmoid curve development, to model the Greek COVID-19 infection curve. The model that we construct is of the form (Norusis, 2008):

$$f(z) = \frac{e^z}{1 + e^z} = \frac{1}{1 + e^{-z}}$$

(14),

where $z$ is the predictor (independent) variable and $f(z)$ is the response (dependent) variable. More broadly, variable $z$ represents a group of independent variables, whereas $f(z)$ determines the probability of the outcome caused due to this group (Walpole et al., 2012). Variable $z$ also expresses the total contribution of all independent variables in the model and is defined as (Norusis, 2008):

$$z = \beta_0 + \beta_1 X_1 + \beta_2 X_2 + \cdots + +\beta_k X_k \qquad (15),$$

where $\beta_0$ is the intersect term of the regression line that is equal to $z$ when all independent variables are zero, whereas $\beta_i$ are the regression coefficients expressing the contribution of each variable to the total variation of the model. A positive value of coefficient indicates that the predictor variable increases the probability of a successful outcome (i.e. the event occurs) whereas a negative value implies that the variable reduces the probability of this outcome. High coefficient values express a strong influence of the independent variable to the formation of the dependent variable, whereas low values indicate small contribution of the independent variable on the occurrence of the corresponding outcome (Norusis, 2008).

To calculate the upper bound of the model forecasting, a hybrid method was used based on averaging. At first, the highest value (i.e. 3,435 infected cases) that was predicted by the Regression Splines algorithm (for the 127th day) was considered. The second value was calculated according to a heuristic approximation based on the formula:

$$U_{p_{b2}} = f(x_{127}) * \langle m \rangle = 3,435 * 1,049521 = 3,932 \qquad (16),$$

where $f(x_{127})$ is the number of total infections in Greece at the 127th day (equal to 3,435 infected cases) and $\langle m \rangle$ is the average of the infections' change rate, which was computed according to the relation:

$$U_{p_b} = \frac{U_{p_{b1}} + U_{p_{b2}}}{2} = \frac{3,435 + 3,932}{2} = 3,683 \qquad (17),$$

By applying this approach for estimating the maximum outcome of the logistic regression, we obtained the results shown in Table 2.

**Table 2**
Logistic Model Summary and Parameter Estimates

| Equation | Model Summary | | | | | Parameter Estimates | |
|----------|----------|---|-----|-----|------|----------|-----|
|          | R Square | F | df1 | df2 | Sig. | Constant | b1  |
| Logistic | 0.938 | 790.679 | 1 | 52 | 0.000 | 0,137 | 0.853 |

According to Table 2, there is a considerably high level of determination of the model ($R^2$=0.938), which allows predicting future values of the Greek pandemic curve in high precision. This determination level appears particularly important because it provides an effective model for predicting the flattening of the pandemic curve in Greece and is crucial for providing insights about the effectiveness at which the Greek health system is expected to respond in the future provided that the pandemic will be described by the same dynamics. Such information will be directly linked to the application of extra or the relaxation of social measures.

## 6. Discussion

The previous analysis illustrates that the proposed approach facilitated the construction of a reliable model since in all cases it led to accurate results that strengthened the forecasting process. In addition, one of the major advantages of the model regards its high reliability due to low residuals, which can be considered as a result of the added value obtained from the complex network conceptualization of the splines regression algorithm.

The complex-network-defined splines regression algorithm provides a solid background delimiting the most relevant dataset in the forecasting process. It is also notable that the proposed methodology models the spread of the Greek pandemic in the most realistic way because, except from the infection data, it takes into account the real context of the applied policies, which improves the level of complexity in the proposed methodology and therefore its realism.

Finally, the added value of the regression splines technique applied in this paper should be evaluated in conjunction with the improvements that the proposed method provides in terms of high accuracy, low multi-collinearity, and model specification. The proposed model offers high accuracy in prediction and stability, as its overall behavior is less noisy, while reducing the overall risk of a particularly poor choice cause from insufficient sampling or model configuration. This is also supported by the variance of the expected error that is minimized, implying the reliability and the universality of the proposed model.

However, an important restriction affecting the reliability of the proposed methodology is the availability of the time-series data used for the model construction. Within this context, the complex network conceptualization of the spline regression algorithm equips the model with additional degrees of freedom to the extent that it incorporates latent information that becomes visible or applicable only through a higher order transformation (i.e. from time-series to a complex network) applied to the source time-series.

Within this context, the ability to accurately predict the future of a pandemic is an extremely important but difficult task. Due to the restricted current knowledge of the new COVID-19 pandemic, to the high level of uncertainty, and to the complex socio-political factors that influence the spread of the new virus, any technically sound methodology for analyzing or predicting the phenomenon is an important legacy and hopeful supply in the fight against the disease.

## 7. Conclusions

This paper studied the COVID-19 temporal spread in Greece and proposed an innovative, realistic, and highly reliable methodology for forecasting the flattening of the curve, based on the spline and logistic regression algorithm, along with the complex network analysis. A considerable added value of the proposed model concerns the complex-network-based conceptualization of the time-series partition into spline segments, according to which the knot-vector of the spline algorithm was defined by the community detection analysis applied to the graph associated to the time-series of the COVID-19 infection curve. In general, the proposed model can create highly realistic scenarios for the evaluation and study of the pandemic, which are directly related to the sociability and mobility of citizens. The very high determination achieved by the proposed model is indicative to its effectiveness and reliability to the extent that it incorporates fitting techniques of high resolution with latent information being visible after transforming the time-series into a complex network. Provided that this study aimed to develop a framework for selecting the appropriate research methods for the further comprehension and interpretation of the disease, as well as to facilitate critical decisions regarding the allocation and management of the available health resources, the proposed methodology contributed to this purpose by introducing an integrated framework incorporating several modeling techniques of time-series analysis. The particularly low error rates that resulted by the proposed method contribute to decision-making and to the examination of the pandemic at a wider spatio-temporal level.

The overall approach showed that, up to the day that Greece opened its borders to foreign tourism, the COVID-19 infection curve was moving into a saturation point describing the flattening dynamics of the curve. This seemed to allow the country deciding to be receptive to the externality of foreign tourism, which started at the very next day of the end of the dataset considered in this study. Within this context, this paper introduces avenues for further research both in methodological and in implementation terms. In methodological terms, future improvements of the proposed methodology should focus on further optimization of the algorithm parameters that used in the model, both of the regression splines and logistic regression algorithms and of the associated complex network conceptualization. Mode avenues of further research could emerge towards the direction of future model expansion, such as the implementation of a hybrid learning system based on the proposed architecture, which with methods of self-improvement and redefinition of its parameters can be lead to a fully automated forecasting. In implementation terms, further research should build on the current dataset of the COVID-19 infection time-series curve in Greece, aiming to measure the effect of foreign tourism in the evolution of the disease, as well as on incorporating more national cases into a broader spatio-temporal study of pan-European or world level, so that to verify the generalization of the method in more complex environments.

Overall, this paper provided a modeling and forecasting tool facilitating decision making and resource management in epidemiology, which can contribute to the ongoing fight against the pandemic of COVID-19.

**Conflicts of Interest:**

The authors declare no conflict of interest.